\documentclass[doublecol]{epl2}
\usepackage{epsfig,graphicx}%[draft]
\usepackage{dcolumn}%Align table columns on decimal point
\usepackage{bm}% bold math
\usepackage{amsmath}
\usepackage{ulem}
\usepackage{color} %to be able to strikethrough text

\title{A dipolar self-induced bosonic Josephson junction}
\shorttitle{A dipolar self-induced bosonic Josephson junction} %Insert here a short version of the title if it exceeds 70 characters

\author{M. Abad\inst{1} \and M. Guilleumas\inst{1} \and R. Mayol\inst{1} \and M. Pi\inst{1} \and D. M. Jezek\inst{2}}
\shortauthor{M. Abad \etal}

\institute{
  \inst{1} Departament d'Estructura i Constituents de la Mat\`{e}ria,
Facultat de F\'{\i}sica, Universitat de Barcelona - E--08028 Barcelona, Spain\\
  \inst{2} IFIBA-CONICET and Departamento de F\'{\i}sica, FCEN-UBA - Pabell\'{o}n 1, Ciudad Universitaria,
1428 Buenos Aires, Argentina
}

\abstract{
We propose a new scheme for observing Josephson oscillations and macroscopic quantum
self-trapping in a toroidally confined Bose-Einstein condensate: a dipolar self-induced Josephson junction.
Polarizing the atoms perpendicularly to the trap symmetry axis, an effective
ring-shaped, double-well potential is achieved which is induced by the dipolar interaction.
By numerically solving the three-dimensional time-dependent Gross-Pitaevskii
equation
we show that coherent tunneling phenomena such as Josephson oscillations and quantum
self-trapping can take place. The dynamics in the self-induced junction can be qualitatively described by a two-mode model taking into account both $s$-wave and dipolar interactions.
}

\begin{document}

\maketitle

\section{Introduction}
Josephson effects are a signature of quantum coherence in macroscopic many-body systems.
Firstly predicted and observed when two superconductors were connected
through a weak link \cite{Tinkham}, Josephson effects have also been experimentally observed in a variety of systems: in superfluid helium
flowing through a sub-micrometer aperture \cite{Avenel1985} and through an array of nano-apertures \cite{Pereverzev1997, Hoskinson2005}; in contact-interacting Bose-Einstein
condensates confined in a double-well trap~\cite{Albiez2005, Levy2007} and in an optical lattice \cite{Cataliotti};   and very recently in exciton-polariton systems in semiconductors~\cite{Lagoudakis2010}.
All these systems are realizations of a Josephson junction (JJ).
Internal Josephson dynamics  has been also experimentally observed between different hyperfine states of
a spinor Bose-Einstein condensate in
Ref.~\cite{Zibold2010}.
In Bose-Einstein condensates (BECs), due
to the nonlinearity introduced by the s-wave contact interaction, there appears
a new phenomenon called macroscopic quantum self-trapping \cite{Smerzi1997, Raghavan1999, Albiez2005}, characterized by
the locking of most of the atoms in one of the two wells.

Dipolar Bose-Einstein condensates
(dBECs)~\cite{Griesmaier2005, Beaufils2008} offer a new playground for the study of
Josephson effects due to the anisotropic and long-range
character of their interaction~\cite{Lahaye2009a}.
Here we present a novel scenario for investigating the coherence properties of quantum transport:
a dipolar self-induced Josephson junction (SIJJ).
Such a junction is not directly created by an external potential, but rather based on the appearance of an effective ring-shaped double-well potential in a toroidally confined dipolar condensate due to the anisotropy of the interaction.
Josephson dynamics in dipolar condensates \cite{Xiong2009, Asad2009} and in spinor dipolar condensates \cite{Tsubota} has been addressed in the literature for external double well potentials, but this is the first time to our knowledge that a SIJJ is investigated.

\section{System description}
We consider $N=5\times10^4$ atoms of $^{52}$Cr with magnetic dipole moment $\mu=6\ \mu_B$ (being $\mu_B$ the Bohr magneton) confined in a pancake-shaped toroidal trap~\cite{Abad2010}
\begin{equation}
\label{Vtrap}
V_{\text{t}}(\mathbf{r})= \frac{m}{2}(\omega_{\perp} r_\perp^2+\omega_z z^2)
   + V_0 \, \exp ( -2 \, r_\perp^2/\; \sigma_0^2) \,,
\end{equation}
where $m$ is the atomic mass, $r_\perp=\sqrt{x^2+y^2}$ is the distance to the trap symmetry axis, and
$\omega_{\perp}=8.4\times2\pi$~s$^{-1}$ and $\omega_z=92.5\times2\pi$~s$^{-1}$ are the radial and axial harmonic trap frequencies. $ V_0=30\ \hbar\omega_\perp$ and
$\sigma_0=2.1\ a_\perp$
%$\sigma_0=10\ \mu$m
are the strength and waist of the Gaussian beam that creates a hole in the condensate along the $z$ axis,
with $a_\perp=\sqrt{\hbar/(m\omega_\perp)}$ the radial oscillator length.
We consider the dipoles oriented perpendicularly to the trap axis (say $y$ axis), and a small and repulsive $s$-wave scattering length, $a=14\ a_B$ (with $a_B$ the Bohr radius),
to ensure that the dBEC is stable but dominated by the dipolar interaction.
In this configuration the axial symmetry imposed by the external confinement is broken, and the anisotropic character of the dipolar interaction is enhanced \cite{Abad2010}.
This effect has been also addressed in dipolar Fermi gases confined in very narrow rings \cite{Dutta2006}, and recently, in bosonic and fermionic dipoles in a quasi-one-dimensional ring trap \cite{Zollner2010}.

\begin{figure}[h]
\epsfig{file=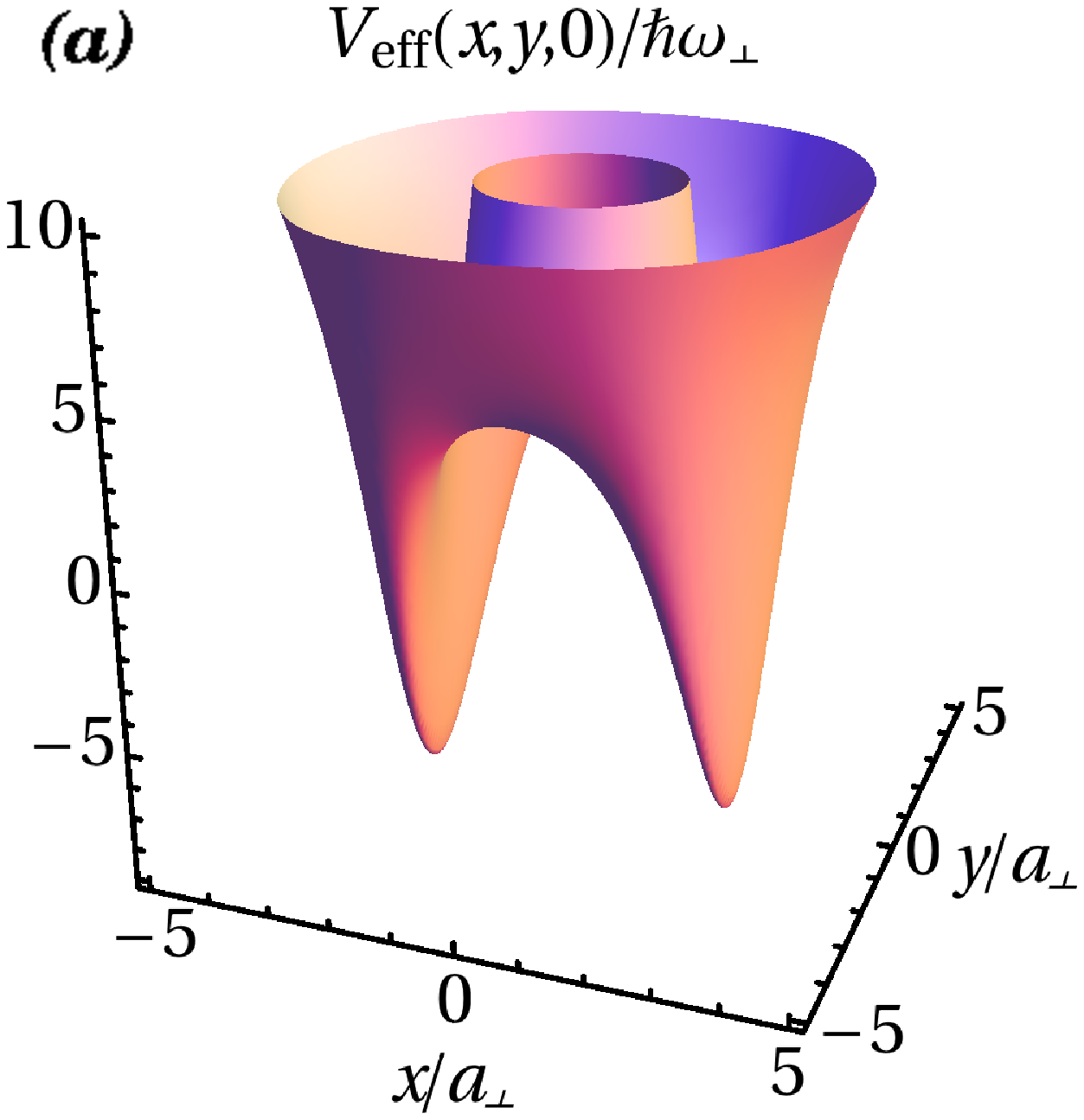, clip=true, width=0.5\linewidth}\epsfig{file=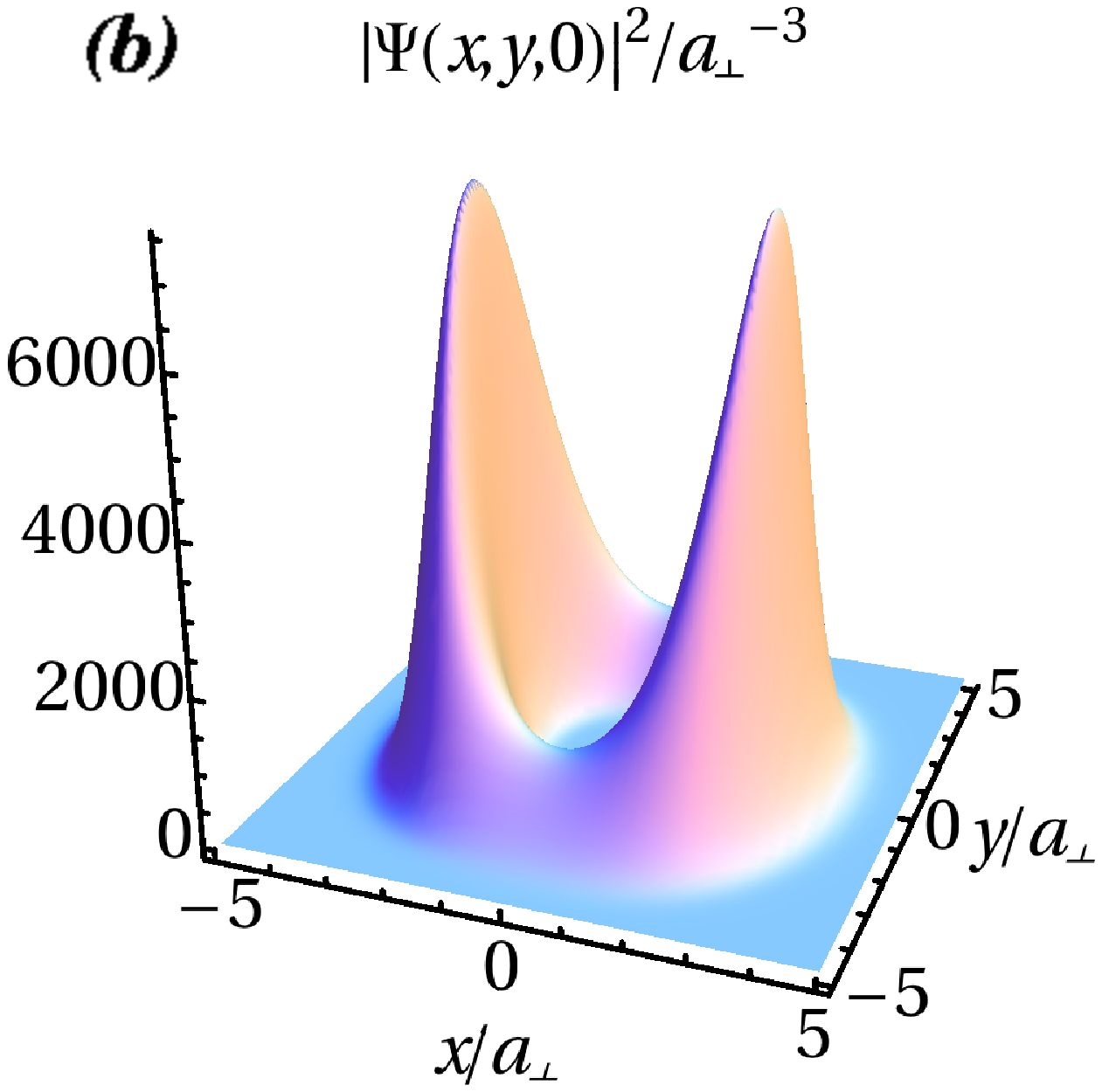, clip=true, width=0.5\linewidth}\vspace{2em}
\epsfig{file=fig1c.eps, clip=true, width=0.465\linewidth}\hspace{0.04\linewidth}\epsfig{file=fig1d.eps, clip=true, width=0.495\linewidth}
\caption{Effective potential (a) and corresponding ground state density (b) in the $z=0$ plane.
Minimum effective potential (c) and maximum density (d) in the $z=0$ plane as a function of the azimuthal angle $\varphi$.
}
\label{dispot}
\end{figure}

The combination of the external toroidal trapping and the mean-field dipolar potential, $V_{\text{d}}(\mathbf{r})$, defines an effective potential,
$V_{\text{eff}}(\mathbf{r})= V_{\text{t}}(\mathbf{r})+V_{\text{d}}(\mathbf{r})$, which has the shape of a ring-shaped double well.
This structure is shown in fig.~\ref{dispot} (a) for the $z=0$ plane, with the two potential wells in the direction perpendicular both to the trap symmetry axis and the polarization axis.
Since the centered Gaussian potential introduces a strong repulsive barrier at $r_\perp=0$ that prevents the atoms from tunneling through it, the double well structure arises in the azimuthal direction. Figure~\ref{dispot}~(c) shows the minimum effective potential along the azimuthal coordinate, $\varphi$, with the two minima at $\varphi=0$ and $\varphi=\pi$ and the two barriers at $\varphi=\pi/2$ and $\varphi=3\pi/2$.
Note that it is thus not defined at a fixed $r_\perp$, but $r_\perp$ slightly changes with $\varphi$, being however close to $r_\perp\sim2.5\,a_\perp$.
According to this effective ring-shaped double well, the atoms localize mostly in the attractive regions inside the wells, producing an azimuthal density dependence. This is seen in panels (b) and (d) of fig.~\ref{dispot}, which show the density and the maximum density along $\varphi$ on the $z=0$ plane, respectively.

The system
resembles two dipolar condensates in a ring-shaped geometry that are coupled via two links. If the barrier height, $\Delta V_\text{eff}$, is large compared to the chemical potential, $\tilde{\mu}$, the system behaves as two weakly linked condensates.
The weak link condition can be reached by tuning the $s$-wave scattering length to small values, which should still be large enough to prevent spatial spontaneous symmetry breaking \cite{Abad2010}.
We find that the ground state fulfills $\Delta V_\text{eff}/\tilde{\mu}=1.1$, which is at the limit of the weak link condition.
The system can thus be thought of as a self-induced Josephson junction. In the remaining of this letter we will show that
it can indeed behave as a bosonic JJ, presenting Josephson as well as self-trapping dynamics.
Note here that this junction really consists of two coupled SIJJs, which in this scenario behave in phase in much the same way as the array of nano-apertures does in the experiments with helium \cite{Pereverzev1997, Hoskinson2005}.

For the sake of completeness we will briefly describe the system for scattering lengths below $a=14\, a_B$. For values still close to this one, the effect of a reduction of the scattering length is translated into an increase of the ratio $\Delta V_\text{eff}/\tilde{\mu}$, but with the effective potential still having the shape of a double well. However, reducing further the scattering length to about $a=12\, a_B$ 
%(just above collapse) 
a spatial symmetry breaking occurs. All the atoms occupy only one of the wells, and the system resembles an origin-displaced cigar-shaped condensate \cite{Abad2010}. Symmetry breaking phenomena in atomic condensates confined in double-well traps have been widely addressed in the literature, both for contact interacting BECs \cite{Mayteevarunyoo2008,Julia2010} and dBECs \cite{Xiong2009,Asad2009}.
 %in a number of systems: in double-well \cite{Cirac1998,Mayteevarunyoo2008,Xiong2009,Asad2009,Julia2010} and double-ring \cite{Malet2010} potentials, in the process of vortex nucleation \cite{Dagnino2009}, and also in dBECs confined in triple-well potentials \cite{Lahaye2010}.
%Such spontaneous symmetry breaking phenomena 
They are thought to be a signature of a quantum phase transition, during which the state of the system is strongly correlated and entangled, forming a macroscopic Schr\"{o}dinger-cat state \cite{Julia2010,Cirac1998,Carr2010}.
%Recently, there has also been some effort to relate self-trapping phenomena in double well potentials to the existence of (dynamical) cat states \cite{Carr2010}. 
For dipolar condensates, such strongly delocalized states have also been theoretically proposed in a three-well configuration \cite{Lahaye2010}.

%Symmetry breaking phenomena are a signature of a quantum phase transition and are presently under strong discussion \cite{Cirac1998,Dagnino2009,Julia2010}.%
%In the last years, symmetry breaking phenomena have attracted a lot of interest. They have been predicted to occur in $s$-wave condensates, for example in double-well potentials \cite{Mayteevarunyoo2008} and double-rings \cite{Malet2010}. And also in dipolar condensates, both in double-well \cite{Xiong2009,Asad2009} and toroidal \cite{Abad2010} configurations.
%Symmetry breaking phenomena are a signature of a quantum phase transition. The state of the system in the transition is strongly entangled, a macroscopic Schr\"{o}dinger-cat state \cite{Cirac1998}.
%Recently, there has been some effort to relate symmetry breaking phenomena in cold atoms to the existence of a cat state \cite{Dagnino2009,Julia2010}. For dipolar condensates, such a strongly entangled state a Schrodinger-cat state has been theoretically proposed in a three-well configuration \cite{Lahaye2009b}.}

\section{Dynamics in the SIJJ}
We study the dynamics of this dipolar SIJJ within the mean-field framework
by solving the full-3D time-dependent Gross-Pitaevskii (TDGP) equation:
\begin{eqnarray}
&&\hspace{-0.5cm} \left[ -\frac{ \hbar^2}{2m} \nabla^2 + V_{\text{t}}(\mathbf{r})
+ \, g|\Psi(\mathbf{r},t)|^2 +  V_{\text{d}}(\mathbf{r},t) \right]\!\! \Psi(\mathbf{r},t)= \nonumber\\
&& =\text{i}\hbar\frac{\partial}{\partial t} \Psi(\mathbf{r},t) \,,
\label{tdgp}
\end{eqnarray}
where $g=4 \pi \hbar^2 a/m$ is the coupling constant of the contact interaction. Note that the mean field dipolar potential, $V_{\text{d}}(\mathbf{r},t)$, depends on time
through the wave function $\Psi(\mathbf{r},t)$. At each time step, it can be
 written in terms of the microscopic dipole-dipole interaction as
\begin{equation}
  V_{\text{d}}(\mathbf{r},t) =\frac{\mu_0 \mu^2}{4 \pi}
\int \text{d}\mathbf{r^\prime}  \frac{1 - 3 \cos^2 \theta}{|\mathbf{r}-\mathbf{r'}|^3}   |\Psi(\mathbf{r^\prime},t)|^2 \,,
\end{equation}
 where $\mathbf{r}$ and $\mathbf{r^\prime}$ are the positions of two aligned dipoles and
$\theta$ is the angle between their relative position
and the polarization axis.
The TDGP equation with the dipolar term becomes an integro-differential equation.
To solve the time evolution
we have used a Hammings algorithm (predictor, corrector, modifier) initialized
by a fourth-order Runge-Kutta method. The dipolar term has been treated using Fourier transform techniques (see Ref.~\cite{Abad2009} and references therein).
The spatial grid used in the computation contains $128\times128\times64$ points with a spacing of
%$0.6$, $0.6$ and $0.4\ \mu$m,
$0.12$, $0.12$ and $0.08\ a_{\perp}$,
respectively for the directions $x$, $y$ and $z$.
The time step used is
%$3\times 10^{-6}$ s.
$1.6\times 10^{-4}\, \omega_{\perp}^{-1}$.

Josephson effects in BECs are characterized in terms of two conjugate variables: the population imbalance and the phase difference between the two wells
\cite{Albiez2005,Smerzi1997,Raghavan1999}. From fig.~\ref{dispot}
we define the left and right wells as the regions where $x <0$ and $x>0$, respectively.
The population imbalance and phase difference are given
by $ Z(t)= (N_L(t)-N_R(t)) / N$ and $ \phi(t)=\phi_R(t)-\phi_L(t)$, respectively,
where $N_{L(R)}(t)$ corresponds to the number of atoms on the left (right) well,
and $\phi_{L(R)}(t)$ the phase of the dipolar condensate averaged on the left (right) well.
To prepare the system with an initial population imbalance  $ Z(0) \ne 0 $, we solve the stationary Gross-Pitaevskii equation in imaginary time \cite{Abad2010} with a small tilting in the external potential, ensuring thus the desired asymmetry
in the population of the two wells. We then let the system evolve in time without any tilting, according to Eq.~(\ref{tdgp}).

\begin{figure}[h]
\epsfig{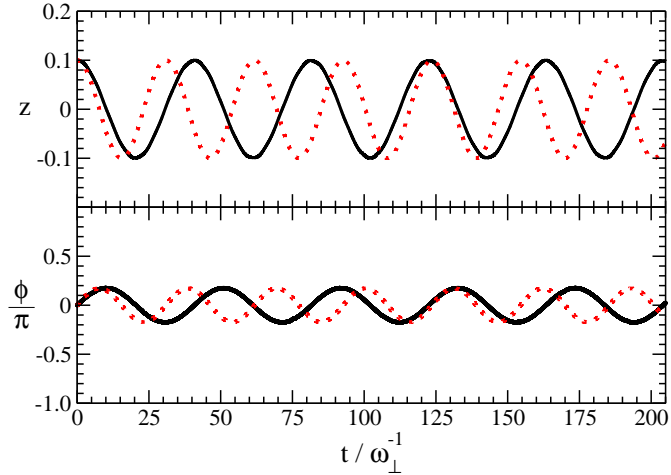}
\caption{ Time evolution of the population imbalance (top panel) and phase difference (bottom panel)
with initial conditions $Z(0)=0.1$ and $\phi(0)=0$.
Solid lines are the TDGP results, and
dashed lines correspond to the TMM.
}
\label{JS}
\end{figure}

For small values of the initial imbalance the system exhibits Josephson oscillations
between the two self-induced wells.
Figure \ref{JS} shows the dynamic evolution of the population imbalance and the phase difference
corresponding to  initial conditions $Z(0)=0.1$ and $ \phi(0) = 0$ (solid lines).
$Z(t)$ and $\phi(t)$ present sinusoidal oscillations shifted by $\pi/2$, and the time average of the population
imbalance is zero.
In this regime, the atoms tunnel periodically from the left to the right well and back. The wave function of the dBEC at each well remains coherent during this process, that is with a uniform phase. This is translated into a phase difference that oscillates in time at the same frequency as the imbalance.
% This point establishes the difference between Josephson oscillations and the more general dipole mode that is defined within the framework of collective oscillations, which corresponds to the center-of-mass motion of the system.

\begin{figure}
\epsfig{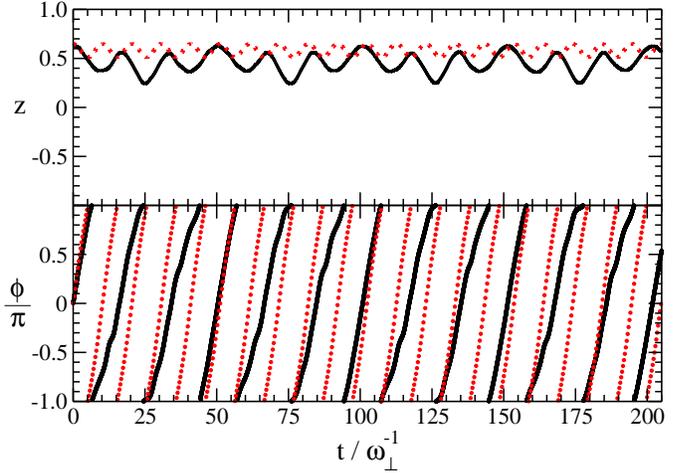}
\caption{ Time evolution of the population imbalance (top panel) and phase difference (bottom panel)
with initial conditions $Z(0)=0.65$ and $\phi(0)=0$.
Solid lines are the TDGP results, and
dashed lines correspond to the TMM.
}
\label{ST_z}
\end{figure}

For a large initial population imbalance, the SIJJ enters the regime of self-trapping oscillations. This can be clearly seen in fig.~\ref{ST_z}, where
the imbalance and the phase
difference are plotted as a function of time for initial conditions $Z(0)=0.65$ and $\phi(0)=0$ (solid lines).
In this situation, the time average of the imbalance remains close to $ 0.5 $ and the phase difference is unbounded (running phase mode). Although in this regime the atoms remain locked in one of the wells, there is still some tunneling of particles at a frequency higher than in the Josephson regime. These imbalance oscillations are non-sinusoidal but have a more involved structure, which hints at a rich self-trapping dynamics.

In external double well potentials, the transition
between Josephson and self-trapping regimes is related to the presence of a saddle point of the Hamiltonian 
in the ($Z,\phi$) space that separates closed (Josephson) and open (self-trapping) trajectories with constant energy \cite{Raghavan1999,Ananikian2006,Mele2010}.
Extending the analogy between external and self-induced double-well potentials, one can expect also in the present case a similar critical behavior separating the two dynamical regimes. 
By solving the time-dependent Gross-Pitaevskii equation, besides the results shown in figs.~\ref{JS} and \ref{ST_z}, we have obtained self-trapping dynamics 
for the initial conditions $Z(0)=0.5$ and $\phi(0)=0$, 
which is compatible with the
critical imbalance obtained with a two mode model, $Z_c=0.43$ when $\phi(0)=0$ (see
below).
%
%Thus, the 
%critical population imbalance that gives the onset of self-trapping dynamics must be smaller than $0.5$.}
%
%In the framework of two-mode models (see below) it is possible to derive an analytical expression for the %critical population imbalance above which the system shows self-trapping. For the parameters used in the %present work, the prediction for the critical imbalance is $Z_c=0.43$, which is in agreement with the results %obtained by solving the time-dependent Gross-Pitaevskii equation (see figs.~\ref{JS} and \ref{ST_z}).}
%as well as with the result (not presented) for $Z(0)=0.5$, which shows self-trapping.}

Figures~\ref{JS} and \ref{ST_z} show that a toroidally confined dBEC behaves as a SIJJ, and that such a structure can sustain both Josephson and self-trapping dynamics. In this sense, the SIJJ is robust, since the dipolar interaction is strong enough to keep the double-well shape of the effective potential in time, even when the initial imbalance is large. We have therefore shown that a new class of systems exists where the double-well potential structure is self-induced by an anisotropic interaction (in this case the dipolar interaction), and that for these systems Josephson and self-trapping oscillations are predicted to occur depending on the initial population imbalance between the two wells.

\section{Two-mode model}
To gain insight into the tunneling dynamics obtained by evolving the TDGP equation, we
have performed a two-mode analysis of the SIJJ, taking into account both s-wave and dipolar interactions.
The two-mode model (TMM) lies on the assumption that the dynamical behavior of a JJ can be fully captured by analyzing the coherent dynamics between two spatially localized modes: the left and right modes, respectively $\Phi_L(\mathbf{r})$ and $\Phi_R(\mathbf{r})$. In this approximation, the condensate order parameter is written using the ansatz~\cite{Smerzi1997, Raghavan1999}
\begin{equation}
 \Psi(\mathbf{r},t) = \psi_L(t)\Phi_L(\mathbf{r}) + \psi_R(t)\Phi_R(\mathbf{r}) \,,
 \label{twomode}
\end{equation}
with $\langle \Phi_i|\Phi_j \rangle=\delta_{ij}$, and the coefficients fulfill  $\psi_j(t) = \sqrt{N_j(t)}e^{i\phi_j(t)}$,
$i,j=L,R$. Note that in this ansatz the time evolution is contained only in the coefficients $\psi_j(t)$.
%Although the Gross-Pitaevskii equation is a nonlinear equation and therefore the superposition principle does in general not hold, ansatz (\ref{twomode}) has proven to be a good approximation provided that the overlap between the two modes is small (condition of weak link) \cite{Albiez2005, Ananikian2006}.

By substituting ansatz (\ref{twomode}) into eq.~(\ref{tdgp}) and performing some algebra
retaining all the overlaping terms~\cite{Giovanazzi2000, Ananikian2006}, one obtains the two-mode equations for a symmetric dipolar SIJJ,
\begin{align}
 \dot{Z} = & (-1+\alpha)\sqrt{1-Z^2}\sin\phi + \varepsilon(1-Z^2)\sin2\phi \label{SIimbalance}\\
 \dot{\phi} = & \Lambda Z -(-1+\alpha)\frac{Z}{\sqrt{1-Z^2}}\cos\phi- \varepsilon Z\cos2\phi\ , \label{SIphase}
\end{align}
with
\begin{align}
 \Lambda = & \frac{U}{2K} - \frac{B+ 2I_1 + D_1}{2K} \label{lambda}\\
 \alpha  = & \frac{I_{2}+D_{3}}{K}\,N \label{alpha}\\
 \varepsilon  = & \frac{I_1+D_1}{2K}\,N \ .\label{varepsilon}
\end{align}
The different parameters appearing in (\ref{lambda})--(\ref{varepsilon}) are given by the integrals
\begin{align}
  K =& -\int \text{d}\mathbf{r}  \left(-\frac{\hbar^2}{2m}\nabla\Phi_L\nabla\Phi_R + \Phi_L\Phi_RV_{\text{t}}\right) \label{k}\\
  U =& \int \text{d}\mathbf{r} \left( g|\Phi_L|^4 + |\Phi_L|^2\int \text{d}\mathbf{r}^{\prime} v_\text{d}|\Phi_L^\prime|^2  \right) \label{u}\\
  B =& \int \text{d}\mathbf{r}\, |\Phi_L|^2\int  \text{d}\mathbf{r}^\prime v_\text{d}|\Phi_R^\prime|^2 \label{b}\\
 I_1 =& \int \text{d}\mathbf{r}\, g|\Phi_L|^2|\Phi_R|^2 \label{i1}\\
 I_2 =& \int \text{d}\mathbf{r} \,g|\Phi_L|^2\Phi_L\Phi_R  \label{i2}\\
 D_1 =& \int  \text{d}\mathbf{r}\,\Phi_L\Phi_R\int \text{d}\mathbf{r^\prime} v_{\text{d}} \Phi_L^\prime\Phi_R^\prime \label{d1}\\
 D_{3}=& \int  \text{d}\mathbf{r}\,\Phi_L\Phi_R\int \text{d}\mathbf{r^\prime} v_{\text{d}} |\Phi_L^\prime|^2 \label{d3} \ ,
\end{align}
where
\begin{equation}
v_{\text{d}}=\frac{\mu_0 \mu^2}{4\pi}\frac{1 - 3 \cos^2 \theta}{|\mathbf{r}-\mathbf{r'}|^3}
\end{equation}
is the microscopic dipole-dipole interaction and $\Phi^{\prime}_{L(R)}\equiv\Phi_{L(R)}(\mathbf{r^{\prime}})$.
Note that the only parameters that appear in eqs.~(\ref{SIimbalance}) and (\ref{SIphase}) are $\Lambda$, $\alpha$ and $\varepsilon$, which depend on both dipolar and $s$-wave interactions.
%, in contrast to the TMM derivations found in the literature \cite{Smerzi1997, Raghavan1999, Giovanazzi2000, Ananikian2006, Xiong2009, Asad2009}.
%Recently, the effects of the competition between long-range and local interactions in the stationary solutions of a one-dimensional double-well potential have been addressed in Ref.~\cite{Wang2010} (XECCCC!!!).
To compute the integrals (\ref{k})--(\ref{d3}) and therefore to determine the parameters (\ref{lambda})--(\ref{varepsilon}), we have used the modes defined as \cite{Ananikian2006}: $\Phi_{L(R)}(\mathbf{r})=(\Phi_{\text{s}}(\mathbf{r})\pm\Phi_{\text{as}}(\mathbf{r}))/\sqrt2$, where $\Phi_{\text{s}}(\mathbf{r})$ and $\Phi_{\text{as}}(\mathbf{r})$ are the symmetric (ground state) and antisymmetric (first excited state) wave functions of the double well potential of fig.~\ref{dispot}. For the case analyzed here, $\Lambda=17.84$, $\alpha=0.1307$ and $\varepsilon=0.1156$.

In the framework of the two-mode model it is possible to find an analytical expression for the critical population imbalance that gives the onset of self-trapping (see, for instance, refs.~\cite{Raghavan1999,Ananikian2006}). For $\phi(0)=0$, it reads:
\begin{equation} 
Z_c=2\frac{\sqrt{1-\alpha}\sqrt{\Lambda-\varepsilon-1+\alpha}}{\Lambda-\varepsilon} \,,\end{equation}
which yields $Z_c=0.43$ in the present configuration.

We have obtained the dynamics within the TMM by solving eqs.~(\ref{SIimbalance}) and (\ref{SIphase}). The results are given as dotted lines in figs.~\ref{JS} and \ref{ST_z}, for the same initial conditions
as the corresponding TDGP calculation.
We see from the figures that the TMM is a good qualitative approximation to the full dynamics given by the TDGP eq.~(\ref{tdgp}):
the order of magnitude of the frequency and amplitude of the oscillations, as well as the dynamical regime imposed by the initial conditions, are well predicted.

The discrepancy between the TDGP dynamics and the TMM can be attributed to a combination of different factors. Firstly, it has been recently shown that the fact that the system does not lie in the deep weak-link limit gives rise to a frequency in the two-mode approximation larger than the experimental one~\cite{LeBlanc2010}, which lies closer to the TDGP result. Secondly, the self-induced nature of the double well means that it depends on time, which is not taken into account in a TMM with time-independent parameters.
Lastly, the SIJJ presented here is clearly two-dimensional, whereas the TMM mimics it as one-dimensional. Dynamics in other directions different than $x$ might
 affect the behavior of the imbalance and the phase difference \cite{Smerzi2003,Mele2010}.
This opens the possibility of new physical aspects of a SIJJ which are not present in a usual JJ.

\section{Extension to other systems}
The whole system is scalable in terms of the dimensionless constants $\lambda=\omega_z/\omega_\perp$, $\tilde{\sigma}_0=\sigma_0/a_\perp$, $\tilde{C}=4\pi Na/a_{\perp}$, $D=N\mu_0\mu^2m/(4\pi\hbar^2a_\perp)$ and $\tilde{V}_0=V_0/(\hbar\omega_\perp)$, which are the coefficients of the different terms of the dimensionless TDGP equation.
For the case considered here $\lambda=11$, $\tilde{\sigma}_0=2.08$, $\tilde{C}=96.77$, $D=24.99$ and $\tilde{V}_0=30$. With these dimensionless constants it would be easy to export the same physics to another set of parameters experimentally accessible in $^{52}$Cr \cite{Griesmaier2005, Beaufils2008, Koch2008}, or even to condensates of alkali gases such as $^{39}$K or $^{7}$Li, where dipolar effects have been observed in Refs.~\cite{Fattori2008} and \cite{Pollack2009}, respectively, making use of a soft zero crossing of the scattering length near a Feshbach resonance.
To give an example of an atomic species with a smaller dipolar moment, we consider a condensate
of $^{39}$K ($\mu\sim1\,\mu_B$) with $10^6$ atoms. The self-induced double well configuration could be obtained in a toroidal trap with $\omega_\perp=65\times2\pi$ s$^{-1}$, $\omega_z=715\times2\pi$ s$^{-1}$ and $\sigma_0=4\,\mu$m, which can be achieved experimentally, 
and with a scattering length of $a=0.3\,a_B$, which is compatible with the $0.06\,a_B$ resolution reported in ref.\cite{Fattori2008}. The period of Josephson oscillations would be in this case close to 100~ms which is well within the lifetime of the condensate.
Moreover, this scenario can possibly
be extended to other systems, such as exciton-polaritons in semiconductors, which can be polarized and where the ring-shaped geometry is experimentally easier to implement than an external double well \cite{polaritons}.

It is worth stressing that throughout this letter the two junctions have been considered to behave in phase. This is the reason why the whole system can be thought of as a single SIJJ and analyzed in terms of a two-mode model. However, there exists another regime where the two junctions  behave with opposite phases (out-of-phase situation). It corresponds to atoms tunneling through the two junctions with opposite velocities.
% but with a zero total velocity circulation (that is, without creation of vorticity). 
In a more general situation the two phase differences would be independent, opening thus the possibility of establishing the bosonic analog of the superconducting quantum interference device (SQUID) \cite{Tinkham}.
% due to the close analogy between the two systems. 
This analysis is however beyond the scope of the present article.

\section{Conclusions}
In conclusion, we propose a dipolar self-induced bosonic Josephson junction. This junction is created by
the anisotropic character of
the dipolar interaction modulated by a toroidal trap, which gives rise to a ring-shaped, double-well effective potential.
The time-dependent Gross-Pitaevskii equation predicts Josephson oscillations as well as a self-trapping regime in this system, depending on the initial population imbalance. Moreover, the self-induced Josephson junction can be analyzed in a two-mode picture, giving qualitative agreement with the Gross-Pitaevskii results.
%By solving the 3D time-dependent Gross-Pitaevskii equation we investigate the coherent properties of quantum transport, showing that Josephson oscillations and self-trapping dynamics can be sustained.
The physics of the self-induced bosonic Josephson junction is not restricted to chromium condensates, but can also be exported to condensates of alkali atoms with controllable contact interactions as well as to other systems such as exciton-polaritons. We believe that it can also be a starting point to study a bosonic SQUID interferometer.

\acknowledgments

The authors would like to thank B. Juli\'{a}-D\'{\i}az and D. Mateo for helpful discussions.
We acknowledge financial support under Grant No. FIS2008-00421 from MEC (Spain), Grant No. 2009SGR1289 from Generalitat de Catalunya (Spain), and Grant No.
PIP 11420090100243 from CONICET (Argentina).
M. A. is supported by CUR (Generalitat de Catalunya).

\end{document}